\documentclass[12pt]{article}
\usepackage{epsfig}
\usepackage{amssymb,graphicx}
\def\be{\begin{equation}}
\def\ee{\end{equation}}
\def\ba{\begin{array}{c}}
\def\ea{\end{array}}
\def\p{\partial}
\def\ben{\[}
\def\een{\]}

\newcommand{\bea}{\begin{eqnarray}}
\newcommand{\eea}{\end{eqnarray}}

\begin{document}

\titlepage
\vspace*{2cm}

\begin{center}{\Large \bf
Quantum exotic: A repulsive and bottomless confining potential
}\end{center}

\vspace{10mm}

\begin{center}
Miloslav Znojil
\vspace{3mm}

\'{U}stav jadern\'e fyziky AV \v{C}R, 250 68 \v{R}e\v{z}, Czech
Republic\\

e-mail: znojil@ujf.cas.cz

\end{center}

\vspace{5mm}

\section*{Abstract}

On a simple model
$V(x,y)=A\,x^2+B\,y^2+C\,x^2y^2+D\,(x^2y^4+x^4y^2)$ we
demonstrate  that even in a classically repulsive regime (i.e.,
at couplings which make the potential decreasing to $-\infty$ in
some directions) quantum mechanics may still support the purely
discrete spectrum of bound states.  In our example, there exists
a critical boundary of this domain of stability where a further
increase of repulsion causes an explosive escape of particles in
infinity.

\vspace{10mm}

PACS
\hspace{5mm}
03.65.Bz
\hspace{5mm}
03.65.Db
\hspace{5mm}
03.65.Ge
\hspace{5mm}
02.30.Sa

\newpage

\section{Introduction}

Hydrogen atom is one of the best known examples of a confinement
of particles (electrons) in an {attractive} potential.  Its
discrete spectrum does not collapse -- this is not perceived as
as a paradox from the very early days of quantum mechanics
\cite{Messiah}.  The explanation is easily acceptable and goes
back to the uncertainty principle.  The stability of this atom
in the origin may be well extended down to the inverse quadratic
central attraction $V_{(v)}(r) \approx v/r^2$, $r \ll 1$ with a
limited strength, $v > -1/4$ \cite{Readva}. An unprotected fall
of electrons to {\em this} singularity only takes place {\em
beyond} the ``natural" critical coupling $v = -1/4$. Its
existence is not surprising -- one simply re-accepts the safe
classical intuition.

A scarcity of {\em non-central} examples of transition between
confinement and its collapse is surprising and worrying.  In
more dimensions, our intuition may fail. In classical mechanics,
a sign of warning comes from an unexpected emergence of chaos in
the anisotropic Coulomb problem~\cite{Gurzwiller}.  In two
dimensions, the emergence of the classical chaos may serve as a
guide to study of the quantum chaos~\cite{Eckhardt}. This seems
best illustrated by the elementary $\alpha \to 0$ limit of the
quartic polynomial potential $V_{[\alpha]}(x,y) =
x^2y^2+\alpha\,(x^4+y^4)$ which is bounded from
below~\cite{Percival}.

After quantization, the peculiar semi-bounded $\alpha \to 0$
extreme $V_{[0]}(x,y)= x^2y^2$ has re-attracted attention as an
approximate model of a non-abelian field~\cite{Savvidy}.  For
this reason, the mathematical gap has quickly been filled.
Several versions of the rigorous proof of the purely quantum
confinement property at $\alpha=0$ have been delivered by Simon
\cite{Simon}.  A full parallel with the Coulombic stability has
been re-established. On the basis of the Heisenberg uncertainty
principle, each plane wave with energy $E > 0$ which tries to
escape along an axis (say, $x$) in infinity proves unable to do
so due to a decreasing width of its classically permitted narrow
escape corridor $x^2\,y^2 \leq E$ with hyperbolic boundaries.

Many questions arise immediately: What are the limits of
capacity of the narrow tubes to prevent the (classically
permitted) asymptotical ``constant speed" escape of quantum
particles?  What could be a decisive counter-acting mechanism?
An acceleration by repulsion?  Which ``asymptotically
bottomless" repulsive potentials could be interpreted as (say,
two dimensional) asymptotical analogues of the above mentioned
critical attraction $V_{(1/2)}(r) \approx -1/(2r)^2$ ? May a
confinig two-dimensional quantum potential $V(x,y)$ be
asymptotically unbounded from below at all?

In the present note, we intend to provide a few answers which,
in all their incompleteness, do not seem entirely trivial.  Even
for polynomial forces in two dimensions, the abundance of
couplings definitely hinders the classification. The
semi-classical estimates of the number of bound states below a
given energy may become (and often happen to be) meaningless.
Still, we shall keep our mathematics virtually elementary and
emphasize the underlying (and, sometimes, quite unexpected)
physical consequences.

We shall pay attention just to a four-parametric family of
particular sextic polynomial models
\ben
V_{(A,B,C,D)}(x,y)=A\,x^2+B\,y^2+C\,x^2y^2+D\,(x^2y^4+x^4y^2),
\ \ \ \ \ \ \ \ D > 0, \ \ \ C \neq 0.
\een
As we shall see, their special cases with a controllable and
tuneable attraction to infinity may be called repulsive in plain
language.  In a way resembling the studies of the central
attraction $V_{(-1/4\pm \varepsilon)}(r)$ we do not expect any
immediate (and, even less, realistic) applicability of these
repulsive forces. We just seek a connection between unusual
asymptotics and a smooth transition between the confined and
de-confined phase in non-central systems.

\section{Analysis}

\subsection{Spectrum}

In a preparatory step, let us abbreviate $\gamma = \sqrt{D} > 0$
and re-parametrize the couplings $C = 2\gamma\,(\alpha +\beta)$,
$B =\alpha^2-\gamma+\delta$ and $A =\beta^2-\gamma+\delta$.
Conversely, this defines the new parameters in terms of the old
ones,
\be
\alpha = {C \over 4\gamma} + \gamma\, {A-B \over C} , \ \ \ \ \
\beta = {C \over 4\gamma} - \gamma\, {A-B \over C} , \ \ \ \ \
\delta = A + \gamma - \alpha^2.
\ee
Such a change of notation simplifies our following key
observation.

\noindent
{\bf LEMMA}.  The spectrum of energies of the Hamiltonian
\be
H_{(A,B,C,D)}
 = -{\p ^2 \over \p x ^2} -{\p ^2
\over
\p y ^2} + V_{(A,B,C,D)} (x,y)
\label{ponent}
\ee
with the positive parameter $\delta > 0$ is discrete.

\noindent
{\bf Proof}.
Firstly, let us notice that the assumption $C \neq 0$ is purely
technical. Easily, the proof at some $C<0$ would extend up to
$C=0$ since, due to the positive semi-definitness of $x^2y^2$,
we may use the inequality $H_{(A,B,C,D)} \leq
H_{(A,B,C+\varepsilon^2,D)}$.  The discrete spectrum of its
left-hand side implies the discrete form of the spectrum of the
right-hand-side operator.  In the second step, let us pick up a
real (and, temporarily, freely variable) number $M > 1$ and
split our Hamiltonian in two parts,
\ben
H_{(A,B,C,D)} = -{ 1 \over M} {\p ^2 \over \p x ^2} - { 1 \over M}
{\p ^2
\over \p y ^2} -{M- 1 \over M} {\p ^2
\over
\p x ^2} - {M- 1 \over M} {\p ^2
\over \p y ^2} + V_{(A,B,C,D)} (x,y) .
\een
The well known estimate $-{d^2 \over dq^2} + \omega^2 q^2 \geq
|\omega| $ of the harmonic-oscillator Hamiltonian may be
recalled to imply
\ben
{M- 1 \over M}
\left( - {\p ^2 \over \p x ^2}
+ { (\alpha+\gamma y^2)^2 M \over M-1} x^2  \right)
\geq \sqrt{M-1 \over M}\ |\alpha + \gamma y^2|
\een
and
\ben
{M- 1 \over M}
\left( - {\p ^2 \over \p y ^2}
+ { (\beta+\gamma x^2)^2 M \over M-1} y^2  \right)
\geq \sqrt{M-1 \over M}\ |\beta + \gamma x^2|.
\een
We have $|\alpha + \gamma y^2| \geq \alpha + \gamma y^2$ and
$|\beta + \gamma x^2| \geq \beta + \gamma x^2$ so that,
irrespectively of the signs of $\alpha $ and $\beta$, we may
conclude that
\be
H_{(A,B,C,D)} \geq \sqrt{M-1 \over M}\ (\alpha+\beta) -{ 1 \over
M} {\p ^2 \over
\p x ^2} -{ 1 \over M} {\p ^2 \over \p y ^2} +
\left[
\left(
\sqrt{M-1 \over M} -1
\right)
\gamma
+\delta
\right]( x^2 + y^2 ).
\label{nt}
\ee
As long as $\delta > 0$, the new couplings of the quadratic term
remain positive for all the sufficiently large $M>M_{min}$. With
any $M_{min}>1$ such that
\ben
M_{min}+ \sqrt{M_{min}(M_{min}-1)} \geq {\gamma \over \delta}
\een
our Hamiltonian $ H_{(A,B,C,D)} $ becomes minorized by an
ordinary separable harmonic oscillator. We may infer that it
possesses the discrete spectrum only.  QED.

Our LEMMA does not seem surprising.  Indeed, whenever
$\alpha^2+\delta> \gamma$ and $\beta^2+\delta> \gamma$, our
potential $ V_{(A,B,C,D)}(x,y) $ is minorized by its
harmonic-oscillator part.  Abruptly, the situation changes when
we admit the negative values of $A$ or $B$.  The repulsivity
constraint $A < 0$ (i.e., $\gamma -\alpha^2 > \delta > 0$) would
induce an {\em accelerated} escape of a classical particle along
the semi-axes $\pm x$.  For $B<0$ (i.e., $\gamma -\beta^2 >
\delta > 0$) the escape would occur along $\pm y$.  At both
these conditions (i.e., for $\gamma -\max (\alpha^2,\beta^2) >
\delta > 0$), the origin becomes a local maximum of
$V_{(A,B,C,D)}(x,y)$.  Our potential acquires a repulsive and
bottomless form.  At $\alpha=\beta=0$, $\gamma = 1.1$ and
$\delta = 0.1$ its shape is displayed in Figure~1.

\subsection{The ground state energy}

We have to notice that the escape tubes are very deep and not as
narrow as one would expect.  The area of the sections $
V_{(A,B,C,D)}(x,y) =E $ remains infinite (!) at an arbitrary
negative energy $E$. The shape of these sections resembles their
quartic $x^2y^2$ predecessors with a steady narrowing
proportional, say, to $1/x$ for $x\gg 1$.  Still, in contrast to
the positively semi-definite tubes in $V_{[0]}(x,y)\geq 0$,
their present narrowing seems more than compensated by the quick
downward fall of their bottom -- this decrease is proportional
to $-|A|\,x^2$ at $y=0$, i.e. quadratic!  With the same
parameters as above, the situation is illustrated in Figure 2.
In a broad interval of energies $2\sqrt{-E} \in (3,9)$ the
thinning of our escape sinks seems virtually negligible.

A flavour of a paradox strengthens with a subsequent observation
that our only condition $\delta > 0$ of the impenetrability of
sinks in LEMMA is in fact entirely independent of the signs of
$\alpha $ and $\beta$.  A reversal of these signs would change
$C> 0$ into $C < 0$ and flip the {\em quartic}, asymptotically
very strong part of our potential upside down, $(x^2y^2> 0) \to
(-x^2y^2< 0)$. This is a significant change but it only shifts
the energies.  The lower estimate of the ground-state energies
\be
E_{(g)} \geq
\alpha+\beta \equiv
{C \over 2\, \sqrt{D}}
\label{estimaten}
\ee
holds for all the Hamiltonians (\ref{ponent}) with $\delta > 0$.

For a proof, let us replace $M \in (1,\infty)$ by $\varepsilon =
1-\sqrt{1-1/M} \in (0,1)$.  With the above inequality $-{d^2
\over dq^2} + \omega^2 q^2 \geq |\omega|$ applied to eq.
(\ref{nt}) once more, this gives us the $\varepsilon-$dependent
family of estimates
\be
E_{(g)} \geq (\alpha + \beta)(1-\varepsilon) +2\,\sqrt{(\delta
-\gamma\,\varepsilon)(2\varepsilon-\varepsilon^2)}
\label{timaten}
\ee
which confirms eq. (\ref{estimaten}) at any sufficiently small
$\varepsilon$.

An improved estimate of $E{(g)}$ may be computed from eq.
(\ref{timaten}) at a right-hand-side maximum (achieved at an
optimal value $\varepsilon_{(opt)}$). In the most interesting
bottomless case with $\delta <\gamma$ we may denote $\delta
/2\gamma = \rho^2 \in (0, 1/2)$, renormalize $
\varepsilon  = 2 \rho^2\, \eta, \ \eta  \in (0,1)$ and put
\be
E_{(g)} =  (\alpha + \beta) + 2^{3/2}\,\delta
\gamma^{-1/2}\,\max_{\eta \in (0,1)}W(\eta,\theta), \ \ \ \
\theta = -  {\rm Arsinh} \left[
\sqrt{1 \over 2 \gamma}  \ \left( {\alpha +
\beta \over 2}\right) \right]
\label{taten}
\ee
where $W(\eta,\theta)= \eta\, {\rm sinh}\  \theta
+\sqrt{\eta(1-\eta)(1-\rho^2\,\eta)}$ and $\theta \in
(-\infty,\infty)$.

A simplification occurs at $\theta = 0$ where the derivative of
$W(\eta,0)$ with respect to $\eta$ vanishes at a unique root of
an algebraic quadratic equation. We get a unique lower estimate
of energies which is a decreasing function of the parameter
$\rho^2 \in (0,1/2)$,
\be
\max_{\eta \in (0,1)}W(\eta,0)
=\sqrt{ (F(\rho)+\rho^2)(F(\rho)+1) \over (F(\rho)+\rho^2+1)^3 }
\in \left( {1
\over \sqrt{3\sqrt{3}}} ,{1 \over 2}
\right) = (0.43869\ldots, 0.5),
\ee
and
$F(\rho)=\sqrt{1-\rho^2+\rho^4} \in (\sqrt{3}/2,1)$.

At $\theta \neq 0$ a similar formula would contain a root of a
biquadratic equation.  A simple algorithm may be recommended
instead.  Its inspiration comes from an observation that in the
interval $(0,1)$, the graph of the function
$\sqrt{\eta(1-\eta)}$ is just an upper half of a circle.  Its
multiplication by the decreasing function
$\sqrt{1-\rho^2\,\eta}$ only slightly deforms this shape. Its
maximum moves down and to the left.  An addition of a linear
function gives the full graph of $W(\eta,\theta)$ as another
very smooth deformation with the right end shifted up or down.
Our idea is to approximate the decreasing factor
$\sqrt{1-\rho^2\,\eta} \in (1/2,1)$ by a constant.

At an initial $n=0$ and with an extreme choice of $\eta = \eta_n
= 1$ we shall define $ {\rm sinh}\  \theta_n= {\rm sinh}\  \theta /
\sqrt(1-\rho^2\,\eta_n)$ and minorize
\be
W(\eta,\theta)
\geq
W_n(\eta,\theta_n)\times
\sqrt{1-\rho^2\,\eta_n},\ \ \ \ \
\eta \leq \eta_n, \ \ \ \ \
W_n(\eta,\theta_n)= \eta\, {\rm sinh}\  \theta_n
+\sqrt{\eta(1-\eta)} .
\label{minor}
\ee
The (unique eligible) maximum of the simplified function $
W_n(\eta,\theta_n)$ lies at the point $\eta_{n+1} = {\exp
\theta_n /( 2 {\rm cosh}\  \theta_n})$. Its value is easily
found,
\be
\max_{\eta \in (0,1)}W_n(\eta,\theta_n)=
W_n(\eta_{n+1},\theta_n)= { 1+\exp 2\theta_n \over 4 {\rm cosh}\
\theta_n }.
\ee
and remains compatible with the minorization (\ref{minor}). Our
approximate graph over-estimates the correct one for
$\eta>\eta_{n+1}$ and under-estimates it for $\eta <\eta_{n+1}$.
The true maximum must still lie to the left from its guess
$\eta_{n+1}$. The validity of minorization (\ref{minor}) is
preserved at $n+1$.

We may iterate the whole construction until a sufficient
numerical precision is achieved.  Table 1 samples its rate of
convergence for $ {\rm sinh}\ \theta = 1$ and $2$ at
$\rho^2=0.4$.

\begin{table}[t]
\caption{Iterative determination of the lower energy estimates
(\ref{taten})
 (a) for ${\rm sinh}\  \theta= 1$ and
 (b) for ${\rm sinh}\  \theta= 2$.}
 \label{dowe}
\centering
\begin{tabular}{||c||ccc||}
\hline \hline
%
&\multicolumn{3}{c||}{\rm (a)}\\
\hline
{\rm iteration}&$\eta_n$&$\theta_n$&{\rm maximum}\\
\hline 0 & 1.000 & 1.073 & 1.000
\\
1 & 0.895 & 1.0464 & 1.14059
\\
2 & 0.8902 & 1.04521 & 1.141076
\\
3 & 0.889969 & 1.045154 & 1.1410952
\\
4 & 0.8899577 & 1.0451516 & 1.14109612
\\
\hline 0 & 0.889900 & 1.045138 & 1.141101
\\
1 & 0.8899544 & 1.0451508 & 1.14109637
\\
\hline
&\multicolumn{3}{c||}{\rm (b)}\\
\hline 0 & 1.000 & 1.677 & 2.000
\\
1 & 0.9663 & 1.66689 & 2.073945
\\
2 & 0.965570 & 1.6666847 & 2.0739846
\\
3 & 0.9655560 & 1.6666805 & 2.0739853
\\
4 & 0.96555573 & 1.6666804 & 2.0739853
\\
\hline 0 & 0.96555569 & 1.6666804 & 2.0739853
\\
1 & 0.96555572 & 1.6666804 &
2.0739853\\
\hline \hline
\end{tabular}
\end{table}

\section{Summary}

In a weakly anharmonic regime (i.e., say, for $\alpha = {\cal
O}(1) = \beta$ and small $\gamma$ and $\delta$) our estimate
(\ref{estimaten}) looks very perturbative. The ground-state
wavefunctions -- perhaps, variational -- may be expected to lie
very close to the well known harmonic oscillator gaussians.  The
growth of $\gamma$ does not change the picture too much. To our
only surprize, the improved gaussian
\be
\psi(x,y) = \exp\left(-{\alpha \over 2}\,x^2 -
{\beta \over 2}\,y^2 - -{\gamma \over 2}\,x^2\,y^2 \right)
\label{tomate}
\ee
becomes the {\em exact} ground-state wavefunction at $\delta
=0$.

A crisis comes when we try to diminish the coefficients $\alpha$
or $\beta$.  The norm of $\psi(x,y)$ in eq. (\ref{tomate})
starts growing and indicates a possible collapse of the system.
Quickly, we re-establish the positivity of $\delta >0$.  Of no
avail!  The threat of collapse becomes unavoidable.  The
seemingly innocent condition $\delta=0$ acquires its real
physical significance as a point where the quantum
impenetrability of our downward sinks is lost, at
$\alpha=\beta=0$ at least.

A deeper analysis of our LEMMA and its proof at any $\alpha$ and
$\beta$ recovers that after a change of sign of $\delta$, our
estimates start working in an opposite direction.  In
particular, deeply in our escape tubes, the {\em local}
approximants of the bound-state energies move {\em downwards}.
Quantum particles commence an accelerated motion and, after all,
disappear in infinity.  In our bottomless and, now, only a
little bit more repulsive potential, the discrete spectrum of
energies collapses down.

We may conclude that the apparent physical paradox of quantum
confinement in the presence of an overall repulsion is
clarified. It is resolved in full analogy with the central
symmetric attraction $\approx v/r^2$. Beyond certain limit,
the classical picture re-enters the scene.  Nontrivial
mathematics must be used. The present text revitalizes and
generalizes the old Rellich's ideas \cite{Rellich} and their
Simon's ``sliced bread" rediscovery \cite{Simon} to forces which
are {\em not} bounded from below.  In such a case we loose the
safe ``uncertainty principle" intuition (plane waves become
accelerated).  Our ``asymptotically bottomless" forces require a
more tricky treatment (basically, a local harmonic
re-interpretation of transversal modes of the wavefunctions).
Of course, such an analysis may be expected transferrable far
beyond our particular sextic example.

\section*{Acknowledgements}

Support by the grants Nr.  A 104 860 2 (GA AV \v{C}R) and
202/96/0218 (GA \v{C}R) is appreciated.

\section*{Figure captions}

\noindent
Figure 1. The negative half of potential $V_{(-1,-1,0,1.21)}(x,y)$.

\noindent
Figure 2. The energy-dependence of boundaries
$V_{(-1,-1,0,1.21)}(x,y) = E$ at (a) $E = -9/4$, (b) $E =
-25/4$, (c) $E = -49/4$ and (d) $E = -81/4$.


\begin{figure}                    
\begin{center}                         
\epsfig{file=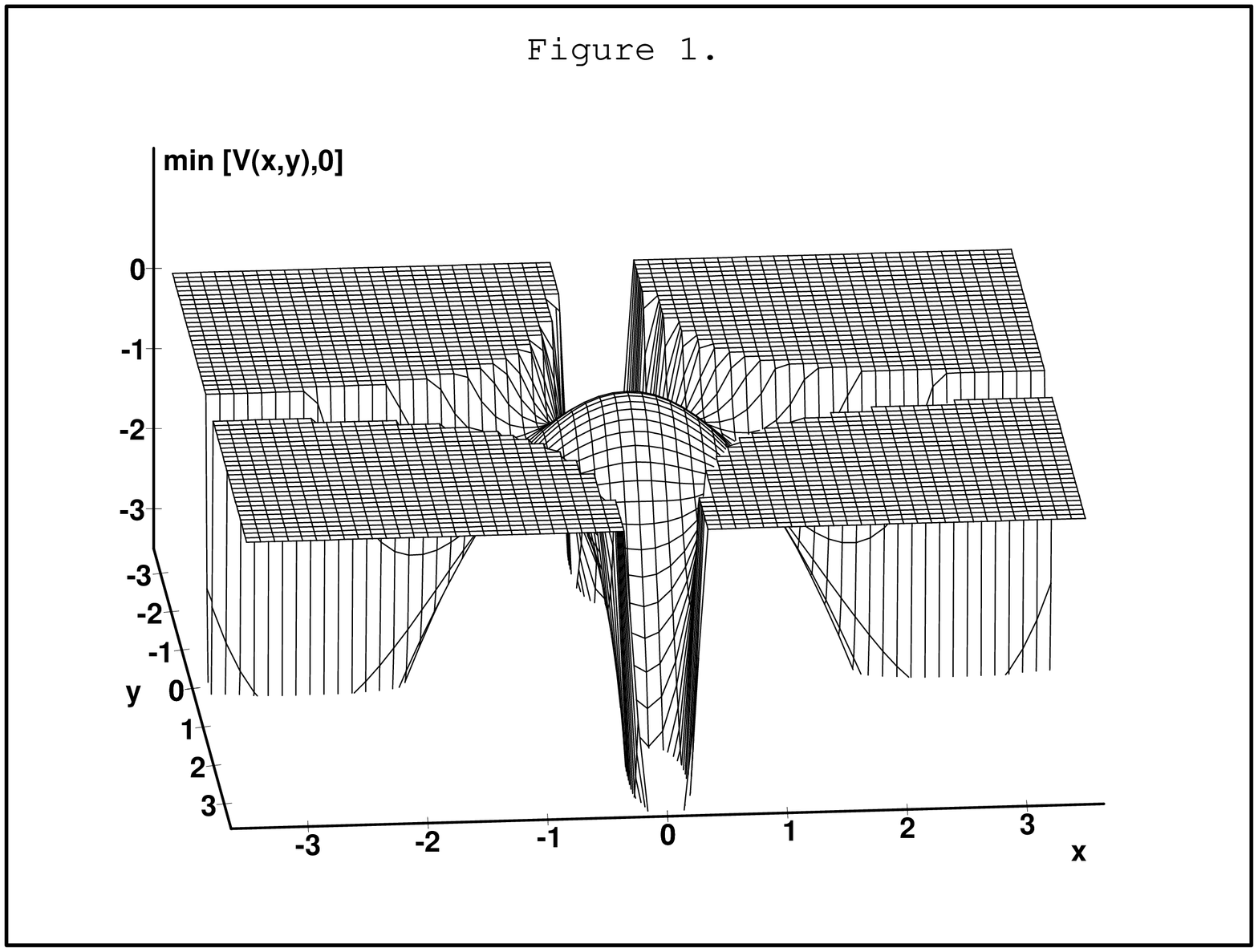,angle=270,width=3.0\textwidth}
\end{center}    
\vspace{2mm} \caption{The negative half of potential
$V_{(-1,-1,0,1.21)}(x,y)$.
 \label{ffit}
 }
\end{figure}
%

\begin{figure}                    
\begin{center}                         
\epsfig{file=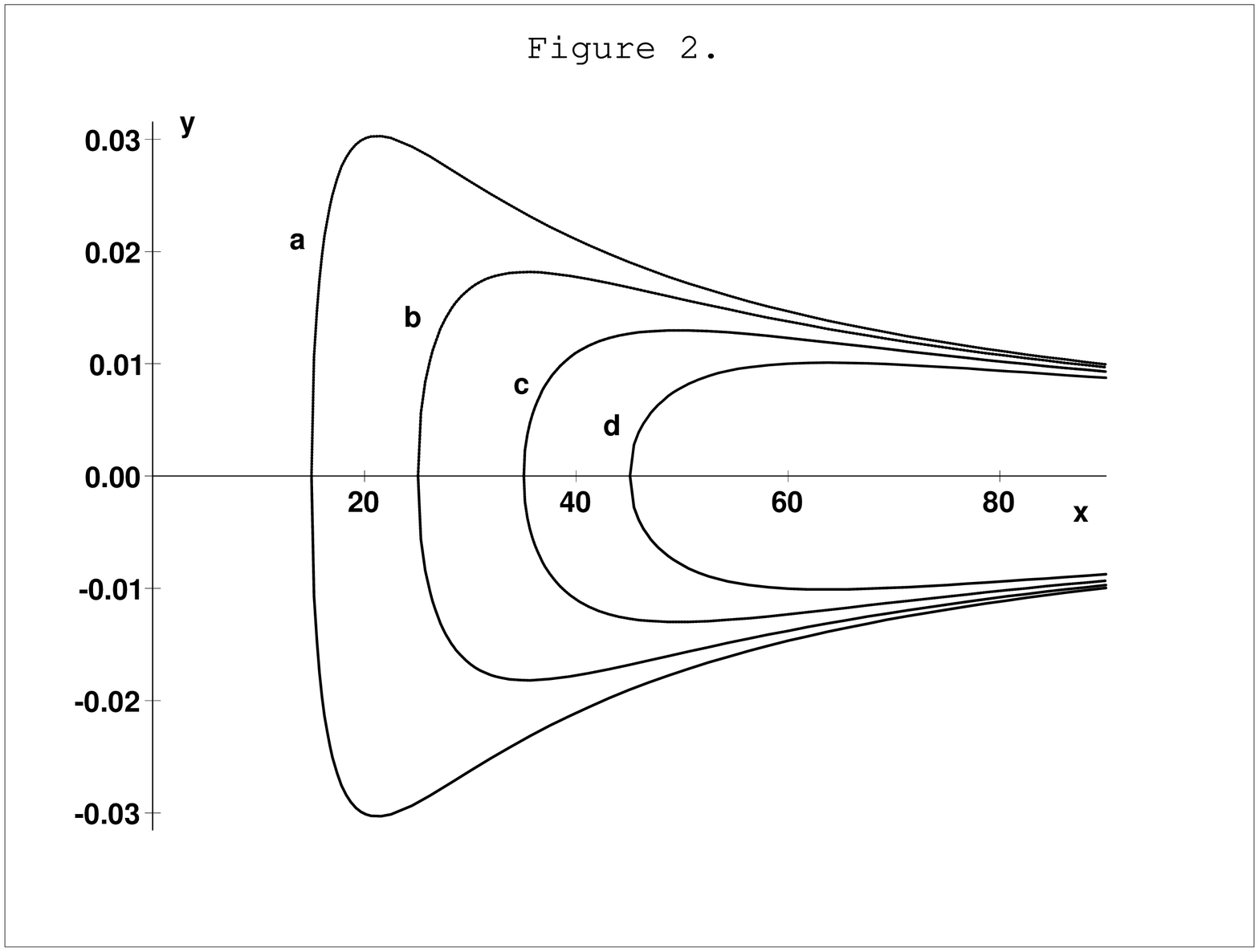,angle=270,width=3.0\textwidth}
\end{center}    
\vspace{2mm} \caption{The energy-dependence of boundaries
$V_{(-1,-1,0,1.21)}(x,y) = E$ at (a) $E = -9/4$, (b) $E = -25/4$,
(c) $E = -49/4$ and (d) $E = -81/4$.
 \label{ffit}
 }
\end{figure}

\end{document}